\documentclass[twocolumn,showpacs,preprintnumbers,amsmath,amssymb,prl]{revtex4}

\usepackage{amssymb}
\usepackage{graphicx}
\usepackage{bm}

\begin{document}
\title{Superconductivity in the topological semimetal YPtBi}

\author{N. P. Butch}
\email{nbutch@umd.edu}
\author{P. Syers}
\author{K. Kirshenbaum}
\author{A. P. Hope}
\author{J. Paglione}
\affiliation{Center for Nanophysics and Advanced Materials, Department of Physics, University of Maryland, College Park, MD 20742}
\date{\today}

\begin{abstract}
The noncentrosymmetric Half Heusler compound YPtBi exhibits superconductivity below a critical temperature $T_c = 0.77$~K with a zero-temperature upper critical field $H_{c2}(0)=1.5$~T. Magnetoresistance and Hall measurements support theoretical predictions that this material is a topologically nontrivial semimetal having a surprisingly low positive charge carrier density of $2\times 10^{18}$~cm$^{-3}$. Unconventional linear magnetoresistance and beating in Shubnikov-de Haas oscillations point to spin-orbit split Fermi surfaces. The sensitivity of magnetoresistance to surface roughness suggests a possible contribution from surface states. The combination of noncentrosymmetry and strong spin-orbit coupling in YPtBi presents a promising platform for the investigation of topological superconductivity.
\end{abstract}

\maketitle

The experimental study of three dimensional topological materials has progressed quickly over the past couple of years. Much recent excitement has surrounded the topological insulator bismuth chalcogenides, in which chiral Dirac-like surface states were predicted theoretically \cite{Zhang09,Xia09} and confirmed by angle-resolved photoemission \cite{Xia09} and scanning tunneling spectroscopy \cite{Cheng10,Hanaguri10}. However, a continuing problem has been that the stoichiometric materials are not bulk insulators as predicted, but rather doped semiconductors \cite{Butch10,Sushkov10,Jenkins10}. This fact has made it difficult to conclusively demonstrate surface transport despite much effort \cite{Checkelsky09,Butch10,Analytis10}. While electrostatic gating techniques show some promise \cite{Cho11,Kim11}, increasing attention is being devoted to other classes of materials \cite{Sato11}.

The rare earth chalcogenide ternary half Heusler compounds, which have a crystal structure that lacks inversion symmetry, were theoretically predicted to harbor topologically nontrivial surface states, depending on the chemical composition \cite{Chadov10,Lin10}. Although there is some debate regarding the topological classification of certain compositions \cite{Vidal11}, YPtBi, LaPtBi, and LuPtBi have a sizable band inversion and are very likely topologically nontrivial. The validity of the band structure calculations has also been called into question by recent angle-resolved photoemission experiments on GdPtBi, DyPtBi, and LuPtBi, which find surface dispersions that differ from those calculated, although the determination of the topology remains unresolved \cite{Liu11}. Despite the recent attention, it is important to remember that the rare earth platinum bismuth series has been previously characterized \cite{Canfield91}, and studied in some detail for its heavy fermion behavior (YbPtBi) \cite{Fisk91} and thermoelectric properties (CePtBi) \cite{Jung01}.

Compared to the other rare earth compounds, YPtBi has received significantly less attention. The initial study by Canfield and coworkers demonstrated that the resistance as a function of temperature had a negative slope between 150~K and 300~K \cite{Canfield91}, while early calculations determined a semiconducting band structure with a narrow 80~meV gap \cite{Eriksson92}. Most experimental characterization has been related to electron spin resonance measurements on lightly doped YPtBi used to study the crystalline electric field states of Nd, Er, and Yb in the nonmagnetic host \cite{Martins95,Pagliuso99}. At low temperatures, the specific heat is dominated by phonons, with a small electronic term $\gamma \leq 0.1$~mJ~mol$^{-1}$~K$^{-2}$, while the magnetic susceptibility is $-10^{-4}$~emu/mol and temperature independent \cite{Pagliuso99}. Here we describe measurements that establish the fidelity of recent band structure calculations identifying YPtBi as a semimetal, so that it can be considered topologically nontrivial, and identify a superconducting ground state. The lack of crystalline inversion symmetry and nontrivial electronic topology raise exciting possibilities for the study of unconventional superconductivity and Majorana fermions.

Single crystals of YPtBi were prepared in excess molten bismuth \cite{Canfield91}. Single crystal x-ray diffractometry showed that the crystal structure is cubic with lattice constant 6.6522(10)~{\AA} at 250~K and space group F-43m (216), as expected for a Half Heusler compound. This lattice constant is in reasonable agreement with the results of recent topological band structure calculations \cite{Feng10,Lin10}. The small diamagnetic susceptibility, indicative of a time-reversal-invariant (nonmagnetic) electronic state was confirmed using a commercial SQUID magnetometer to temperatures as low as 2~K. Between 2~K and 300~K, magnetotransport measurements were performed on a sample sanded into a bar shape (except where noted) in a commercial refrigerator equipped with a 14~T magnet, while transport and ac magnetic susceptibility measurements down to 20~mK were performed in a dilution refrigerator equipped with a 15~T magnet.

\begin{figure}
    {\includegraphics[width=3.1in]{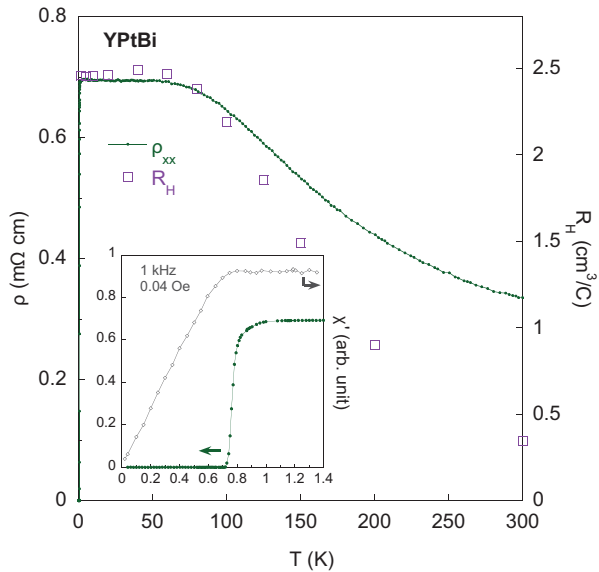}}
    \caption{(color online) The temperature dependence of the resistance of YPtBi is nonmetallic between 0~K and 300~K, probably indicative of a semimetallic band structure. The Hall constant $R_H$ exhibits a similar temperature dependence. Within a one-band model, the carrier concentration varies from $2 \times 10^{18}$~cm$^{-3}$ at 2~K to $2 \times 10^{19}$~cm$^{-3}$ at 300~K. The inset highlights the superconducting transition with critical temperature $T_c = 0.77$~K, as seen in resistivity and ac magnetic susceptibility.}
    \label{RT}
\end{figure}

We first address the bulk band structure of YPtBi. First principles calculations suggest that the compound is semimetallic, but differ in the degree to which the conduction band dips below the Fermi level \cite{AlSawai10,Feng10}. The measured temperature dependence of the electrical resistivity $\rho(T)$ between 20~mK and 300~K is shown in Fig.~\ref{RT}. The behavior is nonmetallic with a negative slope at high temperatures, although it lacks the activated temperature dependence of an insulator.  Below 60~K, $\rho(T)$ is approximately temperature independent, taking a fairly large value of 0.7~m$\Omega$~cm. The overall shape of $\rho(T)$ coincides with that of the positive Hall constant $R_H(T)$, which exhibits a similar plateau in temperature below 60~K. Interpreted within a one-band model, these $R_H$ values correspond to hole densities of $2 \times 10^{18}$~cm$^{-3}$ between 2~K and 60~K.

\begin{figure}
    {\includegraphics[width=3.0in]{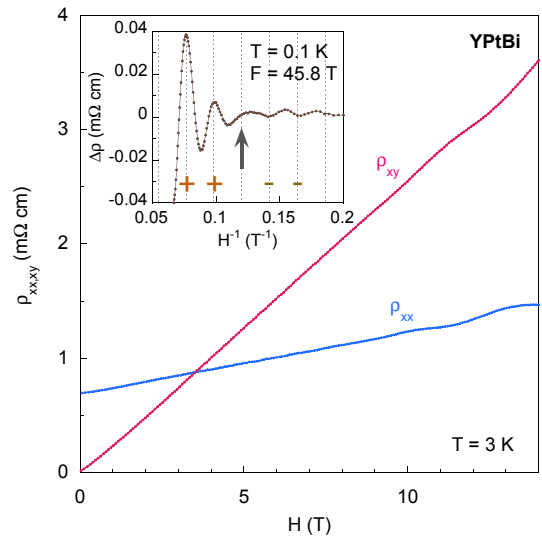}}
    \caption{(color online) Magnetotransport properties of YPtBi with magnetic field along the [001] direction. The low temperature (3~K) transverse magnetoresistance $\rho_{xx}$ and Hall resistance $\rho_{xy}$ exhibit a remarkable linearity up to 14~T. Above 3.5~T, the Hall response exceeds the magnetoresistance, and in both signals, quantum oscillations are clearly visible at higher fields. (Inset) Shubnikov-de Haas oscillations at 0.1~K, with the background magnetoresistance subtracted, have a period of approximately 0.02~T$^{-1}$ (vertical lines) and an effective mass of 0.15~m$_\mathrm{e}$. A node due to beating (arrow) at 0.12~T$^{-1}$ produces a $\pi$ phase shift in the oscillations (+/- signs).}
    \label{MR}
\end{figure}

The low temperature magnetoresistance reveals more about the band structure of YPtBi.  As shown in Fig.~\ref{MR}, the transverse magnetoresistance is unusually linear to magnetic fields of at least 14~T. Shubnikov de Haas oscillations are readily apparent, and can be observed up to temperatures of at least 10~K.  The inset of Fig.~\ref{MR} shows 5 such oscillations measured at 0.1~K, with background subtracted, which have a frequency of approximately 46~T. Assuming a spherical Fermi surface for simplicity, this corresponds to a carrier density of $1.7 \times 10^{18}$~cm$^{-3}$, in good agreement with the Hall measurements. Via a Lifshitz-Kosevich analysis, the temperature dependence of the amplitude of the oscillations yields an effective mass of 0.15 bare electron masses, which is light for a hole pocket. Also of great interest is a node due to beating in the vicinity of 0.12~T$^{-1}$, which is indicative of the presence of two similarly sized spin-orbit split Fermi surfaces. Unfortunately only one node is discernable in our data, otherwise it would be possible to determine the frequencies corresponding to the two Fermi surfaces and estimate the strength of the spin-orbit coupling \cite{Mineev05}. It is also tempting to ascribe the linear magnetoresistance and Hall resistance to Abrikosov's quantum magnetoresistance \cite{Abrikosov00}, although the carrier density $10^{18}$~cm$^{-3}$ is slightly too high for this model to be applicable over the entire field range. Alternatively, the linearity may originate from interband scattering due to the two spin-orbit split Fermi surfaces.

While the low temperature properties of YPtBi are consistent with a spin-orbit split one band description, this breaks down at higher temperatures. For temperatures above 60~K, the single-band Hall carrier density increases by an order of magnitude to $2 \times 10^{19}$~cm$^{-3}$ at 300~K. The Hall mobility $\mu = R_H / \rho_{xx}$ ranges from a respectable value of 3,500~cm$^2$~V$^{-1}$~s$^{-1}$ below 80~K to 1,000~cm$^2$~V$^{-1}$~s$^{-1}$ at 300~K. This behavior implies that most of the reduction in $\rho(T)$ with increasing temperature is nominally due to an unusually large increase in carrier concentration. Strangely, $\mu(T)$ shows a much less dramatic temperature dependence than that of another low carrier density topological metal, namely high-mobility defect-doped Bi$_2$Se$_3$ \cite{Butch10}, begging for an alternative explanation. Consistent with calculated band structures \cite{AlSawai10,Feng10}, a likely scenario is that the Fermi energy falls near the top of a valence band located at the $\Gamma$ point, but less than 10~meV below the bottom of a conduction band close by in the Brillouin zone. As a result, the low temperature properties of YPtBi look like those of a low hole density metal, while increasing temperature above 60~K eventually starts to also populate electron-like carriers, leading to both an increase in scattering and a rapid decrease in $R_H$ due to partial charge compensation. Our transport measurements support the accuracy of the band structure calculations and, as a consequence, nontrivial topology in YPtBi.

\begin{figure}
    {\includegraphics[width=3.0in]{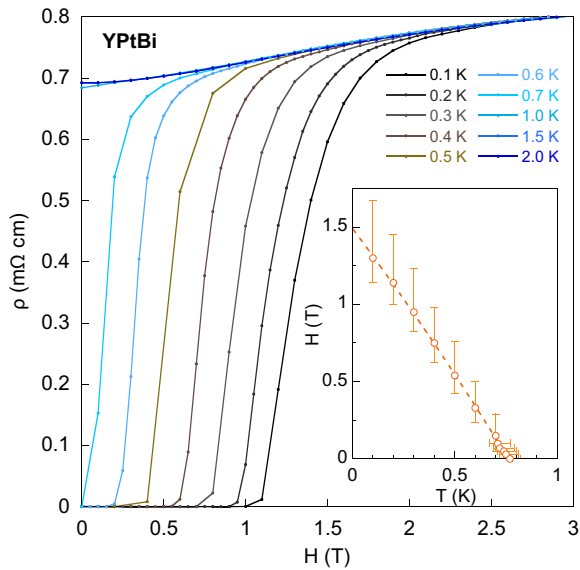}}
    \caption{(color online) Suppression of the superconducting state of YPtBi by magnetic field. The inset shows the unusual quasi-linear temperature dependence of the upper critical field $H_{c2}$ defined as 50\% of the resistive transition. Error bars correspond to 10\% and 90\% of the transition. The zero temperature extrapolated value of $H_{c2}$ is 1.5~T. Such a large $H_{c2}$ value far exceeds the estimate for a conventional orbital limit, but is more compatible with a paramagnetic limit.}
    \label{Hc2}
\end{figure}

Despite the large low temperature resistivity and low carrier density, YPtBi undergoes a superconducting transition at a critical temperature $T_c = 0.77$~K. In fact, YPtBi sits at the extreme of low carrier density superconductors, with even lower carrier concentration than superconducting SrTiO$_{3-x}$ \cite{Schooley65}. The zero field resistive transition, shown in the inset of Fig.~\ref{RT}, is sharp with a gently rolling onset, while a diamagnetic response is observed in the ac magnetic susceptibility. The temperature dependence of the upper critical field $H_{c2}$ with magnetic field applied perpendicular to the [001] direction is shown in Fig.~\ref{Hc2}. The calculated zero temperature orbital limiting field of 0.85~T is significantly lower than the experimentally determined value of 1.5~T. Rather, this value is much closer to the weak-coupling paramagnetic limit of 1.4~T. As shown in the inset of Fig.~\ref{Hc2}, $H_{c2}(T)$ is unusually linear and shows no sign of saturation at low temperatures. The value of 1.5~T corresponds to a superconducting coherence length $\xi = (\frac{\Phi_{0}}{2 \pi H_{c2}})^{\frac{1}{2}}= 15$~nm, where $\Phi_0$ is the magnetic flux quantum, while the low-temperature normal state mean free path $\ell = 130$~nm, so $\frac{\xi}{\ell} \ll 1$ and YPtBi is in in the clean limit.

The isoelectronic compound LaPtBi, with a slightly larger lattice parameter, provides an interesting comparison. Earlier calculations predict a semimetallic band structure \cite{Oguchi01} while recent calculations predict a topological zero-gap semiconductor \cite{AlSawai10,Feng10}. Like YPtBi, single crystals of LaPtBi also exhibit a negative $\rho(T)$ slope, although both the resistivity and carrier density ($4 \times 10^{18}$~cm$^{-3}$) are higher and $\rho(T)$ does not saturate at low temperatures\cite{Jung01,Goll08}. The lack of saturation is a clue to the importance of differences in band structure near the Fermi level. LaPtBi exhibits a relatively linear magnetoresistance, but high-field quantum oscillations suggest that it has a larger Fermi surface \cite{Wosnitza06}, in accordance with the Hall measurements. Finally, LaPtBi also undergoes a superconducting transition at 0.8~K \cite{Goll08}. Yet, given the differences in Fermi surface between LaPtBi and YPtBi, strong spin-orbit coupling, and lack of inversion symmetry, it is plausible that their respective superconducting order parameters are inequivalent.

A strong candidate for topological superconductivity is Cu intercalated Bi$_2$Se$_3$ with $T_c = 3$~K \cite{Hor10}, which maintains distinct surface states despite its high doping level \cite{Wray10}. Recent theoretical calculations suggest that the superconducting state of this highly doped topological insulator may harbor Majorana modes in certain vortices \cite{Hosur11}. The bulk properties of the superconducting state have recently been determined, with some evidence of deviation from the standard Bardeen-Cooper-Schrieffer (BCS) theory \cite{Kriener11} and unconventional tunneling \cite{Sasaki11}. This material is metallic, with a carrier density of $10^{20}$~cm$^{-3}$, two orders of magnitude greater than that of YPtBi, yet the respective values of $T_c$ are fairly similar. It is also noteworthy that YPtBi provides a better crystallographically ordered base for superconductivity, as Bi$_2$Se$_3$ supports a range of Cu concentration that affects $T_c$, and the distribution of Cu may not be homogeneous on small length scales \cite{Kriener11b}.

Although it has not been addressed in detail theoretically, the superconducting state in YPtBi is also a candidate for topological superconductivity on the general grounds presented by Qi and coworkers \cite{Qi10}. Our transport data support strong spin-orbit coupling and the presence of nondegenerate Fermi surfaces, although Coulomb interactions are probably not very strong, based on the measured light carrier masses. Such a topological superconducting state would be unusual, harboring unconventional in-gap bound states \cite{Lu10}. Even in the absence of topological nontriviality, as a noncentrosymmetric superconductor, YPtBi has the potential to harbor mixed singlet and triplet pairing, as is believed to occur in CePt$_3$Si \cite{Scheidt05}.

\begin{figure}
    {\includegraphics[width=3.0in]{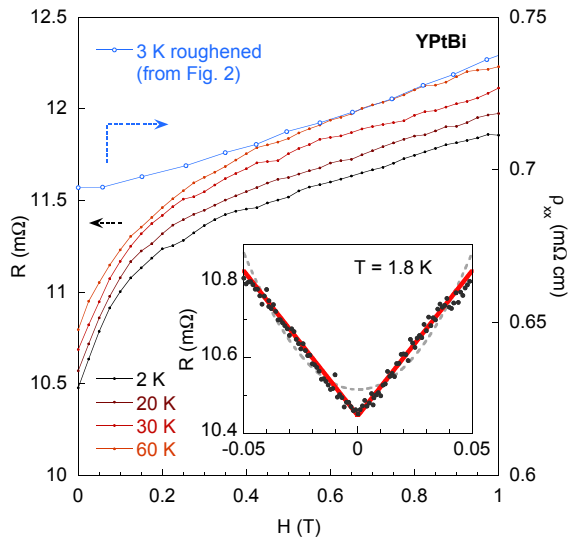}}
    \caption{(color online) Low field transverse magnetoresistance of a YPtBi crystal exhibiting negative curvature. This behavior, which persists to high temperatures, is reminiscent of the weak antilocalization observed in other topologically nontrivial materials. The low field cusp disappears after mechanically roughening the surface, after which the field dependence becomes weakly super-linear. (Inset) The magnetoresistance is unusually linear (solid red line) at very low field, in contrast to conventional quadratic behavior (dashed gray curve). }
    \label{WAL}
\end{figure}

Finally, we highlight an interesting property of YPtBi that is suggestive of behavior observed in some studies of the bismuth chalcogenides. The low-field magnetoresistance of a sample is shown in Fig.~\ref{WAL}, in which a cusp-like field dependence is evident near zero field. The calculated magnetoconductance can be well fit by a logarithmic field dependence similar to that used to describe 2 dimensional weak antilocalization, as has been done for doped Bi$_2$Se$_3$ \cite{Checkelsky09}, thin films \cite{Kim11Oh}, and devices \cite{Kim11}. In this case, the conductance is fit to the form $A [\log(\frac{H_0}{H})-\psi(\frac{1}{2}+\frac{H_0}{H})]$ where $A$ is an amplitude expected to have a value on the order of the conductance quantum, while $H_0$ is a characteristic dephasing field and $\psi$ is the digamma function. This behavior persists over an impressive temperature range, up to at least 60~K before the curve flattens and the magnetoresistance adopts a super-linear field dependence at higher temperatures. Values of $H_0$ range from $8\times10^{-3}$~T at 1.8~K to $2\times10^{-2}$~T at 60~K. Most tantalizingly, the low-field cusp is no longer observed once the sample surface is roughened mechanically (see $\rho_{xx}$ data in Fig.~\ref{MR}). It is tempting to interpret the 2-dimensional fits as evidence for surface state conduction, but the absolute magnitude of the amplitude is several orders of magnitude too large. On the other hand, YPtBi has a fairly high mobility at low temperatures and would seem to be a poor candidate for bulk 3-dimensional weak antilocalization effects.

In conclusion, we have identified superconductivity in the topological low carrier concentration semimetal YPtBi. Strong spin-orbit coupling and the lack of crystalline inversion symmetry make this a promising candidate for the study of mixed pairing and topological superconductivity.

\begin{acknowledgments}
We thank P. Y. Zavalij for crystal structure refinement and K. Behnia, D. Culcer, H. D. Drew, M. S. Fuhrer, and A. H. Nevidomskyy for helpful discussions.  This work was supported by NSF-MRSEC (DMR-0520471), DARPA-MTO (N66001-09-c-2067), and AFOSR-MURI (FA9550-09-1-0603).
\end{acknowledgments}

\end{document}